\documentclass[%
 reprint,
 amsmath,amssymb,
 aps,
]{revtex4-2}

\usepackage{graphicx}% Include figure files
\usepackage{dcolumn}% Align table columns on decimal point
\usepackage{bm}% bold math
\begin{document}

\preprint{APS/123-QED}

\title{GWtuna: Trawling through the data to find Gravitational Waves with Optuna and Jax}% Force line breaks with \\

\author{Susanna Green}
\affiliation{Institute of Cosmology and Gravitation, University of Portsmouth, Portsmouth PO1 3FX, United Kingdom}

\author{Andrew Lundgren}
\affiliation{Catalan Institution for Research and Advanced Studies (ICREA), E-08010 Barcelona, Spain}
\affiliation{Institut de F´ısica d’Altes Energies (IFAE), The Barcelona Institute
of Science and Technology, UAB Campus, E-08193 Barcelona, Spain}
\affiliation{Institute of Cosmology and Gravitation, University of Portsmouth, Portsmouth PO1 3FX, United Kingdom}

\date{\today}

                             %display desire

%\section{abstract}
\begin{abstract}
GWtuna is a fast gravitational-wave search prototype built on Optuna (optimisation software library) and JAX (accelerator-orientated array computation library) \cite{Akiba2019Optuna:Framework, jax2018github}. Using Optuna, we introduce black box optimisation algorithms and evolutionary strategy algorithms to the gravitational-wave community. Tree-structured Parzen Estimator (TPE) and Covariance Matrix Adaption Evolution Strategy (CMA-ES) have been used to create the first template bank free search and used to identify binary neutron star mergers. TPE can identify a binary neutron star merger in 1 second (median value) and less than 1000 matched-filter evaluations when 512 seconds of data is searched over. A stopping algorithm is used to curtail the TPE search if the signal-to-noise ratio (SNR) threshold has been reached, or the SNR has not improved in 500 evaluations. If the SNR threshold is surpassed, CMA-ES is used to recover the SNR and the template parameters in 9,000 matched filter iterations taking 48 seconds (median value). GWtuna showcases alternatives to the standard template bank search and therefore has the potential to revolutionise the future of gravitational-wave data analysis.   
\end{abstract}

\maketitle

\section{Introduction}
In 2017 the first gravitational-wave, GW170817, from
a binary neutron star merger was observed by Advanced
LIGO and Virgo during their second observing run \cite{Abbott2017GW170817:Inspiral}. The binary neutron star merger was also independently detected by Fermi/GBM and INTEGRAL as a short gamma-ray burst and therefore GW170817 marked the dawn of multi-messenger astronomy \cite{Abbott2017Gravitational170817A, Savchenko2017INTEGRALGW170817, Cowperthwaite2017TheTo}. GW170817 led to constraints on the radii of neutron stars whilst the follow-up observations of the event in the optical and infrared spectrum verified that binary neutron star mergers enable the formation of r-process heavy elements \cite{Abbott2018GW170817:State, Cowperthwaite2017TheTo, Coulter2017SwopeSource, Kasen2017OriginEvent, Drout2017LightNucleosynthesis}. Comparison made between the gravitational-wave observation and the electromagnetic counterpart enabled the speed of gravity and light to be constrained to within $10^{-15}$, which helped eliminate alternate theories of gravity used to explain dark energy \cite{Abbott2017Gravitational170817A, Creminelli2017DarkGRB170817A, Ezquiaga2017DarkAhead}. Furthermore, the Hubble constant was also constrained by identifying the host galaxy of the event \cite{Borhanian2020DarkTension, Hotokezaka2019AGW170817, Chen2018AYears}. To date only two binary neutron star mergers have been observed by the LIGO-Virgo-KAGRA (LVK) Collaboration. In the fourth observing run, at least one binary neutron star merger is predicted to be observed and as sensitivity is improved in the fifth observing run in 2027, more are expected to be observed \cite{Abbott2020ProspectsKAGRA}. 

PyCBC \cite{PyCBCSearch2016, Nitz2018RapidLive}, SPIIR \cite{Kovalam2022EarlySearch, Chu2022}, MBTA \cite{Adams2016, Adams2015LowMBTA, Aubin2021}, and GstLAL \cite{Cannon2021GstLAL:Discovery, Ewing2024PerformanceRun, Mukherjee2021TemplateVirgo, Sakon2024TemplateKAGRA} are matched-filter search pipelines used to identify binary neutron star mergers in the LVK Collaboration. All these search pipelines use a pre-generated bank of gravitational-wave templates to extract a gravitational-wave signal from the noisy interferometric data using matched-filtering. Template banks used to identify neutron star mergers in low latency are of the order 10,000 in size \cite{Nitz2018RapidLive, Kovalam2022EarlySearch, Brown2012} which means that for each section of data, $\mathcal{O}(10,000)$ matched-filters are calculated whether a gravitational-wave signal is in the data or not. As a result matched-filtering with template banks results in many unnecessary calculations being performed because binary neutron star mergers are a very infrequent events. There have been attempts to address this issue using alternate methods, such as hierarchical searches, but the sensitivity of the search is compromised \cite{Soni2024ObtainingSearch}. Binary neutron star searches are required to be sensitive and fast because the astronomy community needs to be alerted for electromagnetic follow-up. As much, third generation detectors aim to localise neutron star mergers prior to their merger \cite{Abbott2011ExploringDetectors, Reitze2020CosmicLIGO, Evans2021ACommunity, Punturo2010TheObservatory, Hild2011}.

In this paper, we present a fast gravitational-wave search prototype, called GWtuna, which is built on the Optuna and JAX libraries. Binary neutron star mergers were injected into 512 seconds of coloured noise. The noise was generated with an O4 Advanced LIGO Power Spectral Density (PSD) curve at a sampling rate of 2048 Hz. Tree Parzen Estimator (TPE), a Bayesian optimisation method, was used to identify a peak in SNR by adjusting the mass and aligned spin of the gravitational-wave template in fewer than 1,000 matched-filters. A stopping algorithm is utilised to curtail the search if a gravitational-wave event is not observed in the data or the SNR threshold is surpassed, therefore preventing any unnecessary calculations. If the SNR threshold is reached, then the Covariance Matrix Adaption Evolution Strategy (CMA-ES) is used to recover the SNR and the parameters in 9,000 matched-filters. GWtuna can therefore identify a peak in SNR in less than 10,000 matched-filters, which is an order of magnitude less than standard binary neutron star template bank searches which require 10,000 matched-filters. TPE required 1 second (median value) to identify a peak in SNR whereas, CMA-ES required 48 seconds (median value) to recover the SNR and the parameters. The key functionalities of GWtuna can be summarised: 

\begin{itemize}
  \item Non-template bank search, 
  \item Stopping algorithm, 
  \item JAX functionalities - specifically JIT and GPU compatibility,
  \item Computationally cheap,
  \item Easy to generalise.

\end{itemize} 

We will begin with a brief review of matched-filtering in section \ref{Theory}. Then in section \ref{GWtuna} we will introduce the software, algorithms, and frameworks used in GWtuna, with explanations of TPE and CMA-ES. This is followed by an overview of the GWtuna Search algorithm in section \ref{GWtuna Search}. Section \ref{Results} details the results of GWtuna and how well the SNR is recovered using CMA-ES with population restart and learning rate adaption. We then explore the potential of GWtuna, including a new gravitational wave search pipeline, in section \ref{Discussion}. We then conclude in section \ref{Conclusions} by highlighting the most beneficial functionalities of GWtuna. All the code is open source and can be accessed at https://github.com/SusannaGreen/GWtuna.

\section{Matched-Filtering}
Matched-filtering is a signal processing technique that is used to detect a known signal that is buried in stationary and Gaussian noise. The output of an ideal gravitational-wave detector can be defined as \textit{x(t)=s(t)+n(t)}, where \textit{n(t)} is the noise observed in a gravitational-wave detector and \textit{s(t)} is a gravitational-wave signal. The noise in the detector can be modelled by the one-sided noise power spectral density, $S_{n}(f)$, and is described by  
\begin{equation}\label{average noise}
\langle \tilde{n}^{*}(f) \tilde{n}(f') \rangle = \delta(f-f')\frac{1}{2}{S_{n}(\textit{f})},
\end{equation}
where ${n}^{*}(f)$ is the complex conjugate of \textit{n(f)}, $\delta(f-f')$ is the Kronecker delta function to show frequencies are independent, $\tilde{n}(f)$ is the Fourier transform of \textit{n(f)}, and $\langle \tilde{n}^{*}(f) \tilde{n}(f') \rangle$ is the ensemble average (or mean) of the noise \cite{Allen2012FINDCHIRP:Binaries, Brown2004SearchingHalo}. The Fourier transform of \textit{n(f)} is defined as, 
\begin{equation}
\tilde{n}(f) \equiv \int_{-\infty}^{\infty} \textit{dt} \ e^{-\textit{i}2\pi \textit{ft}} n(t).
\end{equation}
Discrete versions of the integral are used because the output of the gravitational-wave detector is discrete \cite{Allen2012FINDCHIRP:Binaries}. The detector output is a time-series consisting of \textit{N} samples where $N=T/\delta t$, \textit{T} is the total time of the detector output which is sampled every $\delta t$. Therefore, all the possible detector outputs form an \textit{N}-dimensional vector space. Gravitational-wave signals, \textit{s(t)}, can be modelled by a \textit{n}-dimensional parameter space which creates a \textit{n}-dimensional submanifold (\textit{n}-dimensional parameter space in an \textit{N}-dimensional vector space) with the parameters forming a coordinate system \cite{Babak2006}. For compact binary coalescence (eccentricity ignored) the gravitational wave can be parameterised by 15 parameters (the mass and spin angular momentum of the two bodies; the distance to the source; the time of coalescence; the position of the source on the sky which is described by 2 angles; the orientation of the source which is also described by 2 angles, and the orbital phase at a given time) and can be regarded as a 15-dimensional surface embedded in a vector space of all possible measured signals \cite{CurtandFlanagan1994}. Gravitational wave-templates, \textit{h(f)}, are numerical models that describe compact binary mergers and are used in extracting the gravitational-wave signal from the detector output using matched-filtering \cite{Ajith2007AWaveforms}. 

The aim of matched-filtering is to determine which gravitational-wave template produces the highest signal-to-noise ratio (SNR) which is defined as as, 
\begin{equation}\label{SNR}
\rho = \frac{(x|h)}{\sqrt{(h|h)}},    
\end{equation}
where $(x|h)$ is defined as 
\begin{equation}
(x|h) = 4 \Re \int_{0}^{\infty} \frac{\tilde{x}(f)\tilde{h}(f)^{*}}{S_{h}(\textit{f})} \textit{df} 
\end{equation}
which is described as the weighted inner product between the detector output, \textit{x(f)=s(f)+n(f)}, and the gravitational-wave template, \textit{h(f)}. The similarity to Equation \ref{SNR} with the standard definition of a vector, a, with unit norm
\begin{equation}
\hat{a} = \frac{a}{\sqrt{(a|a)}},    
\end{equation}
means that effect of filtering a signal, \textit{s}, with a template, \textit{h}, is equivalent to projecting the signal in the direction of \textit{h} \cite{Babak2006}. In regards to gravitational-wave data analysis, the parameters of the gravitational-wave template that produces the greatest SNR value are not known in advance. Therefore for each section of data the template parameters are varied in order to identify the template that produces the largest SNR. How the template parameters are varied can either be predefined, in the case of template banks, or chosen `on the fly.' In GWtuna, gravitational waves from binary neutron star mergers are searched for using algorithms, specifically TPE and CMA-ES, that enable the template parameters to be chosen `on the fly.' 
\label{Theory}

\section{GWtuna}
GWtuna searches for gravitational waves using software, algorithms, and frameworks that were originally designed for the machine learning community. GWtuna does not use machine learning but the algorithms used in GWtuna `learn' about the mapping between the template parameters and the SNR. Therefore, the template parameters are adjusted throughout a GWtuna search to find the highest valued SNR in the noisy interferometric data. In this section we highlight the software, algorithms, and frameworks used in GWtuna so that they can be fully utilised in the gravitational-wave community. 

\subsection{JAX}
GWtuna is a gravitational-wave search prototype that is built on the JAX and Optuna framework. JAX is a high performance array computing Python library that is able to perform well on large-scale linear algebra and is useful for numerically intensive operations \cite{jax2018github}. JAX was used in GWtuna to enable accelerated linear algebra (XLA) through Just-In-Time (JIT) compilation and GPU compatibility. JIT is used to accelerate the execution of python functions by compiling them into highly optimised code so that it can be run on a GPU. It does this by initially tracing the functions and then building a computational graph that represents the series of operations performed. This then enables an accelerated linear algebra (XLA) compiler to optimise the code fusing multiple operations (multiple mathematical steps, such as multiplication and division, can be done in a single step) and eliminating redundant computations. This optimised graph is then compiled into highly efficient code so that it can be executed on the relevant hardware. JAX can be executed on multiple backends, including GPU, therefore all code was written in JAX NumPy module and ran on an A100 GPU. 

\subsection{Optuna}
Optuna is a hyperparameter optimisation framework and is commonly used in hyperparameter optimisation for various forms of machine learning algorithms \cite{Akiba2019Optuna:Framework}. Hyperparameter optimisation is tasked with finding the parameter values that yield the highest valued objective function. However, identifying these parameter values in machine learning is a challenge for various reasons. One reason is that the objective function is unknown and therefore algorithms are required to learn about the function throughout the optimisation process, which is why it is commonly referred to as black box optimisation. Additional challenges are faced when there are lots of parameters, a noisy objective function, and multi-modality in an objective function. Many of the challenges faced in the machine learning community are also encountered in the gravitational-wave community, specifically with regards to searching for gravitational-wave signals in noisy strain data. In the machine learning community there have been many algorithms have been designed to address these challenges and Optuna contains some of the most popular algorithms, including Tree-structured Parzen Estimator and Covariance Matrix Adaption Evolution Strategy.

\subsection{Tree-structured Parzen Estimator}
Tree-structured Parzen Estimator (TPE) is part of the GWtuna search because it uses Bayesian optimisation to update template parameters and is computationally cheap. TPE was originally proposed in 2011 as a hyperparameter optimisation algorithm for neural networks and deep belief networks \cite{Bergstra2011AlgorithmsOptimization}. Since then TPE has successfully been used in the machine learning community as a Bayesian optimisation method \cite{Watanabe2023Tree-StructuredPerformance}. TPE initially samples random sets of parameters in order to determine which areas of the search space produce the highest valued objective function. Then on each iteration, a Gaussian Mixture model $l(x)$ is fitted to the parameters that produce the best results and another Gaussian mixture model $g(x)$ is fitted to the remaining parameter values. Then the parameters $x$ are chosen so that the ratio $l(x)/g(x)$ is maximised \cite{Bergstra2013MakingArchitectures}. 

\subsection{Covariance Matrix Adaption Evolution Strategy}
The Covariance Matrix Adaption Evolution Strategy (CMA-ES) is one of the most successful black box optimisation algorithms and has been shown to outperform more than 100 optimisation methods \cite{Loshchilov2013BI-populationSearches}. CMA-ES has out performed various algorithms (including Monte Carlo methods) on a variety of problems ranging in dimensionality, multi modality, and noise \cite{Milgo2017ComparisonSamplers, Hansen2009BenchmarkingTestbed, Hansen2004EvaluatingFunctions}. CMA-ES was included in the GWtuna search because previous research has shown that it has performed well on noisy problems and in gravitational-wave astronomy we are searching for a gravitational-wave signal in noisy data. 

CMA-ES is an evolutionary algorithm that is independent of the coordinate system and has been shown to be particularly useful in solving non-separable, ill-conditioned and rugged problems (i.e. sharp bends, discontinuities, outliers, noise, and
local optima) \cite{HansenN2016TheTutorial}. Evolutionary algorithms were originally proposed in the 1970s and their design was based on the biological idea of natural selection. In evolutionary algorithms, steps are chosen at random to search the parameter space in that generation, similar to Monte-Carlo methods. Then the best steps in that generation are used to create the next random steps for the following generation and as a result the most beneficial traits are passed down. Evolution strategy algorithms are one type of evolutionary algorithm that are designed to optimise vectors of real numbers, $x \in \mathbb{R}^{n}$. 

In CMA-ES, samples are initially taken from the distribution. Then parameters of the distribution, specifically the mean and the covariance matrix, are updated based on the objective function value \cite{Hansen1996AdaptingAdaptation}. The mean and covariance matrix are updated to decrease the expected evaluation value which is strongly related to natural gradient descent \cite{Akimoto2010BidirectionalStrategies, Ollivier2017Information-geometricPrinciples}. Natural gradient descent is a second-degree optimisation method that has been motivated to take the steepest descent in the space of the distribution rather than the space of the parameters and the Riemannian metric is used for computing the distance in the distribution  \cite{Shrestha2023NaturalAnalysis, Amari2000MethodsGeometry, Amari1998NaturalLearning}. Research has shown that CMA-ES and other evolutionary strategy algorithms, specifically the Natural Evolution Strategy algorithm, use natural gradient descent (either fully or in part) to search the parameter space \cite{Akimoto2010BidirectionalStrategies, Wierstra2014NaturalStrategies, Sun2009EfficientStrategies}. This ability is one of the reasons why CMA-ES has been so successful.

There are a variety of flavours of CMA-ES that are used to solve a variety of problems (i.e. noisy problems): warm starting CMA-ES \cite{Nomura2021WarmOptimization}, CMA-ES with learning rate adaption \cite{Nomura2023CMA-ESProblems}, CMA-ES with population restart \cite{AugerASize}, CMA-ES with bi-population \cite{Hansen2009BenchmarkingTestbedb}, CMA-ES with margin \cite{Hamano2022CMA-ESMargin}, separable CMA-ES \cite{Ros2008AComplexity}. Research has shown that CMA-ES with population restart and learning rate adaption works well on noisy problems because population restart prevents CMA-ES getting stuck into local minima by increasing the population size on each restart \cite{AugerASize, Nomura2023CMA-ESProblems, Hansen2009BenchmarkingTestbedb}. In GWtuna, CMA-ES with population restart and learning rate adaption is used to search the binary neutron star parameter space to determine the peak in SNR. 

\label{GWtuna}

\section{GWtuna Search}
This GWtuna search was designed to identify binary neutron star mergers by adjusting the template parameters in order to maximize the objective function, which is the signal-to-noise ratio (SNR). In current binary neutron star searches the template parameters that are adjusted are the mass and aligned spin because this is how templates have been designed. However, other parameters can be used to search for binary neutron star mergers. $\eta$ is defined as the mass ratio, $\eta = \frac{m_{1}}{m_{2}}$ where $m_{1}$ and $m_{2}$ are defined as the mass of the two neutron stars. $\mathcal{M}_{c}$ is called the chirp mass and can be expressed as 

\begin{equation}
\mathcal{M}_{c} = \frac{(m_{1}m_{2})^{3/5}}{(m_{1}+m_{2})^{1/5}}. 
\end{equation}

$\lambda_{0}$ has been shown to smooth out the mass parameter space in the binary neutron star region and is determined by the following equation 

\begin{equation}
\lambda_{0} = \left( \frac{\mathcal{M}_{c}}{\mathcal{M}_{c_{ref}}} \right) ^{-5/3}, 
\end{equation}

where $\mathcal{M}_{c_{ref}}$ is a predefined reference chirp mass which for this research was $\mathcal{M}_{c_{ref}}=0.8706$, the lowest chirp mass for a binary neutron star \cite{Brown2012}. In GWtuna, $\eta$ and $\lambda_{0}$ are adjusted in order to identify the gravitational-wave template that produces the largest SNR. The mass range of a single neutron star that is commonly searched over is between $1M_{\odot}$-$2M_{\odot}$, which corresponds to a $\eta$ range of \textit{0.1875}-\textit{0.2499} and a $\mathcal{M}_{c}$ range of \textit{0.8706}-\textit{2.6117} \cite{Abbott2023GWTC-3:Run}. Whilst the spin range that is commonly searched over is between \textit{-0.4} and \textit{0.4} which corresponds to the fastest-spinning pulsar known \cite{Brown2012, Lorimer2008BinaryPulsars}. $\lambda_{0}$, $\eta$ and aligned spins are the parameters that GWtuna, specifically TPE and CMA-ES, are searched over in order to maximise the SNR. GWtuna has the potential to search over as many parameters as desired but there is a computational cost involved with searching over lots of parameters. CMA-ES is a highly parallelisable algorithm but the variation used in Optuna, and therefore GWtuna is not \cite{Nomura2021WarmOptimization}.

In GWtuna, TPE is used to determine whether there is a potential gravitational-wave signal in the strain data. In Optuna, TPE is quicker at searching the $\lambda_{0}$, $\eta$ and aligned spin parameter space compared to CMA-ES. However, CMA-ES is better at recovering the SNR and the parameters compared to TPE. As a consequence, GWtuna is designed so that TPE is used as an initial search algorithm and then CMA-ES is only called if there is a potential gravitational-wave signal in the strain data. TPE has 1000 attempts, or matched-filter evaluations, to maximise the SNR by adjusting $\lambda_{0}$, $\eta$ and aligned spin. If one of the following criterion is met: 

\begin{itemize}
  \item The SNR threshold is not reached and there has been no improvement in SNR after 500 attempts, 
  \item The SNR threshold is reached,
\end{itemize}

the TPE search is curtailed by a stopping algorithm. The first condition is used to prevent unnecessary calculations being performed if there is not a gravitational-wave event in the data. The second condition is used to recover the SNR and parameters using CMA-ES. Once TPE has identified a peak, CMA-ES (with population restart and learning rate adaption) is used to recover the SNR and the parameters in 9000 matched-filter evaluations. CMA-ES with warm-starting was briefly tested for this paper but the results were inferior to the ones shown in this paper.
\label{GWtuna Search}

\section{Results}
\begin{figure}
    \centering
    \includegraphics[scale=0.4]{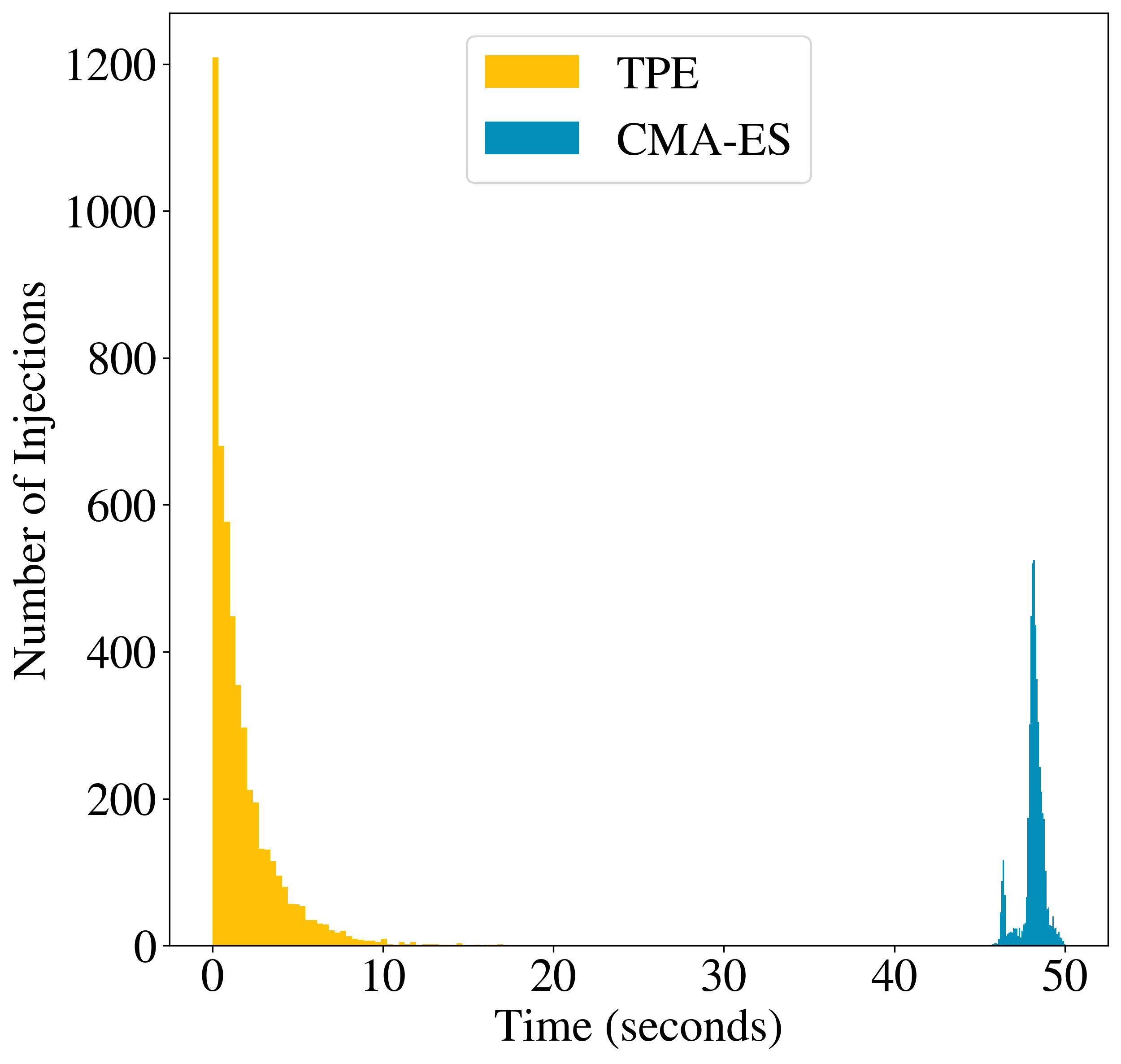}
    \caption{The time taken for TPE and CMA-ES algorithms are compared in this histogram. It takes TPE less than 10 seconds to identify a peak in SNR with a median of 1.09 seconds. Whilst CMA-ES takes on average less than 50 seconds with a median of 47.8 seconds on a GPU A100.}
    \label{fig: TPE and CMAES Time}
\end{figure}

\begin{figure}
    \centering
    \includegraphics[scale=0.4]{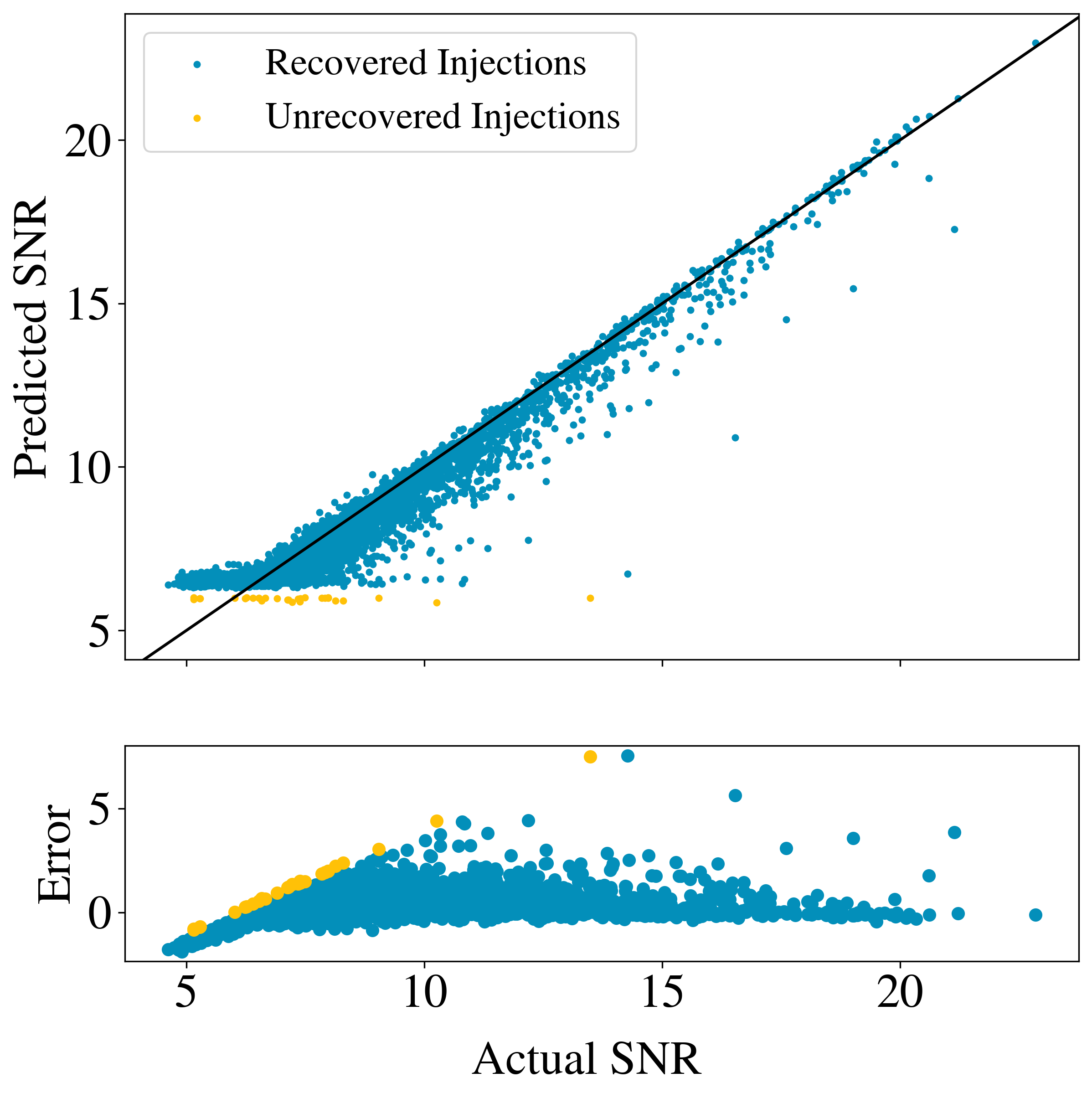}
    \caption{The SNR recovered by the GWtuna algorithm for 5000 injections compared to the actual SNR calculated. The blue dots are used to showcase when TPE has successfully identified a GW in the data whereas the yellow dots are when TPE has been unsuccessful.}
    \label{fig: SNR Actual and Predicted}
\end{figure}

\begin{figure}
    \centering
    \includegraphics[scale=0.3]{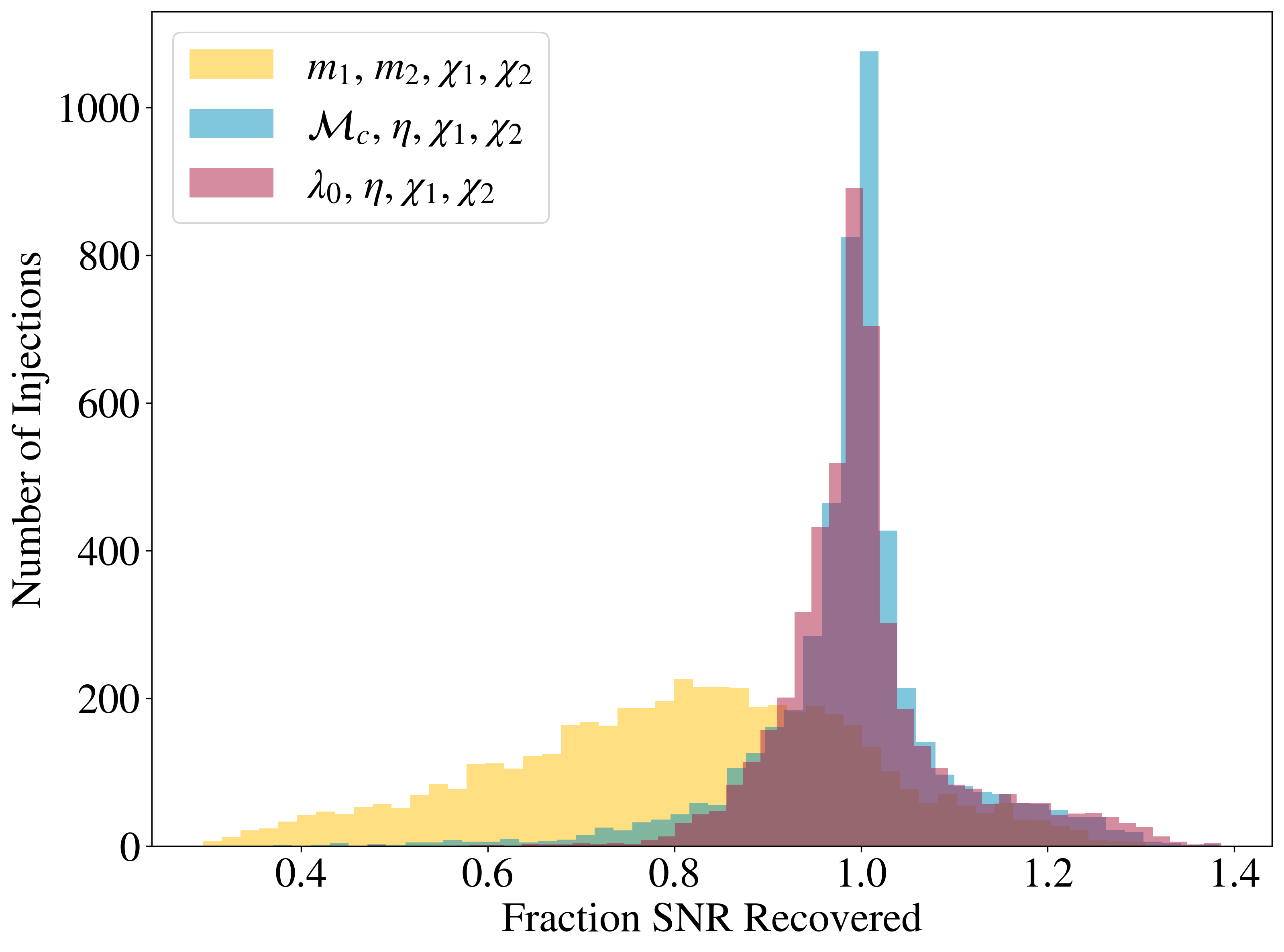}
    \caption{Different variations of mass and spin parameters were tried and it was concluded that $\lambda_{0}$, $\eta$, $\chi_{1}$, and $\chi_{2}$ were the best parameters at recovering the SNR. It is worth noting that $\lambda_{0}$, $\eta$, $\chi_{1}$, and $\chi_{2}$ were generally more precise at recovering the SNR whilst $\mathcal{M}_{c}$, $\eta$, $\chi_{1}$, and $\chi_{2}$ were generally more accurate.}
    \label{fig: Mass Parameter}
\end{figure}

\begin{figure*}
    \centering
    \includegraphics[scale=0.5]{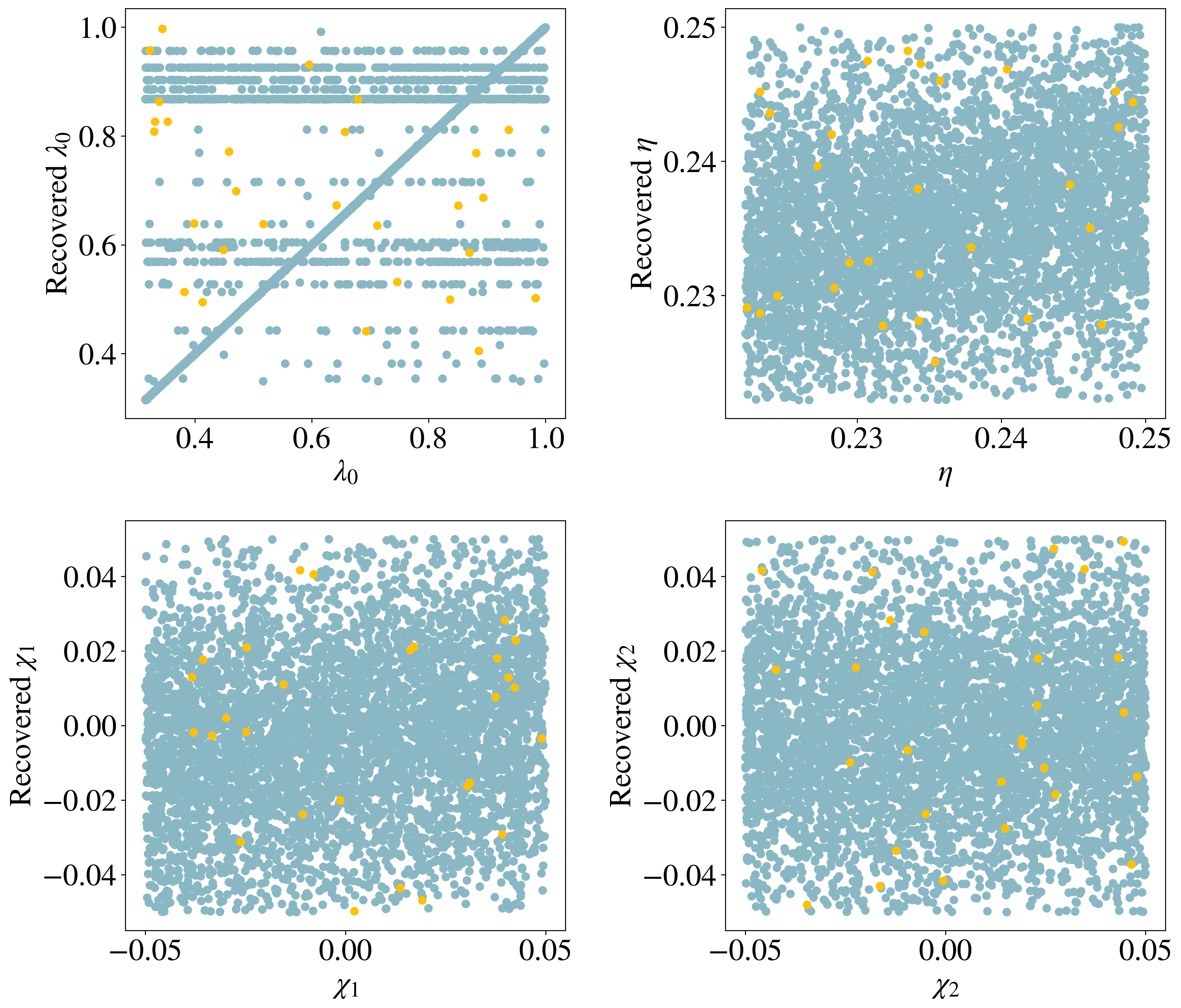}
    \caption{The $\lambda_{0}$, $\eta$, $\chi_{1}$ and $\chi_{2}$ values predicted by CMA-ES compared with actual values. The yellow dots represent the injections that were unsuccessfully recovered by TPE whereas the blue dots were the successful injections.}
    \label{fig: Param recovery}
\end{figure*}

\begin{figure}
    \centering
    \includegraphics[scale=0.3]{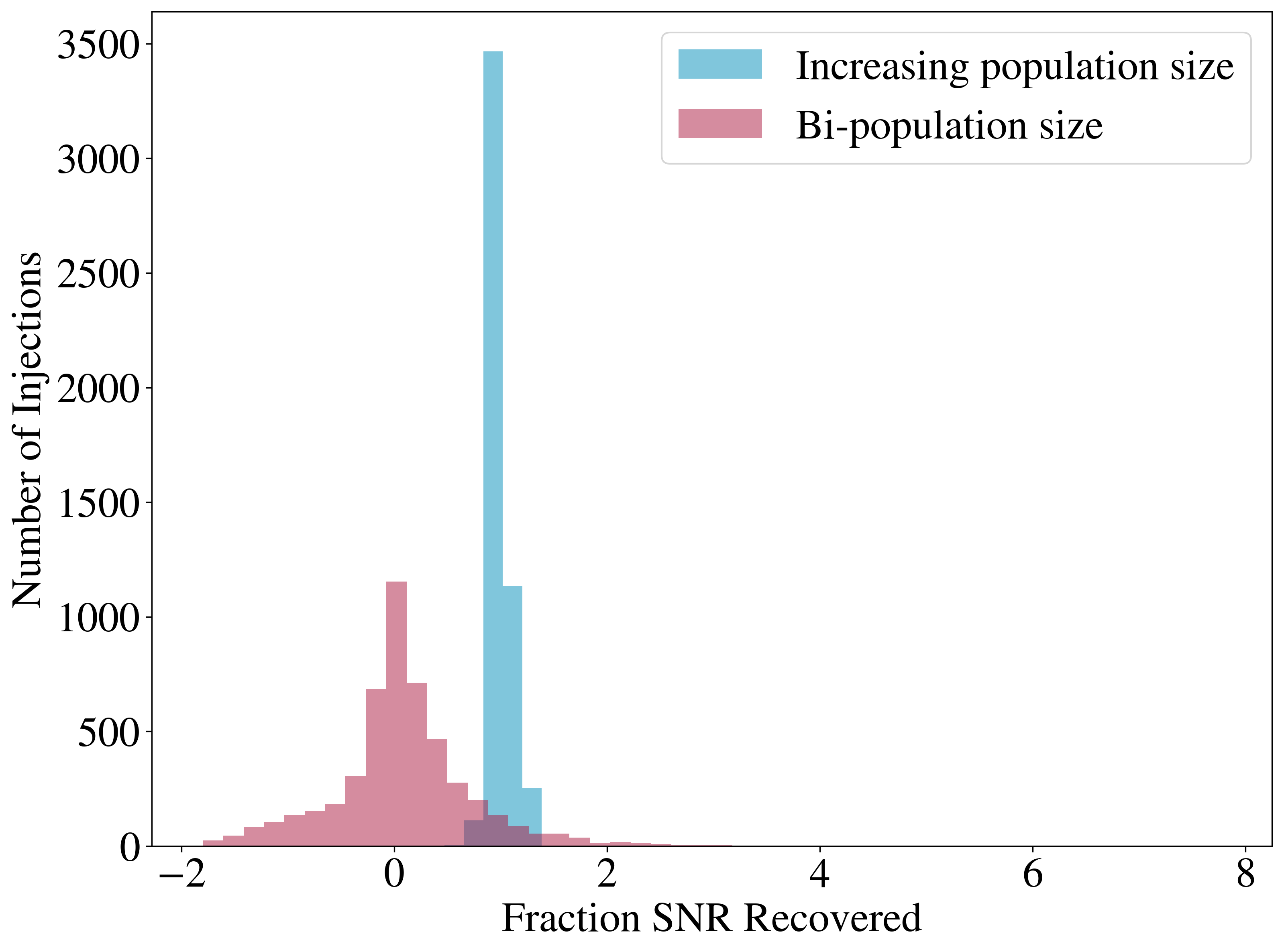}
    \caption{Population size can either be increased or decreased at each restart. In this plot the we compared increasing the population size with increasing or decreasing the population size using the bi-population method. It was concluded that increasing the population size, rather than the bi-population method, recovered the SNR the best.}
    \label{fig: ipop or bipop}
\end{figure}

\begin{figure}
    \centering
    \includegraphics[scale=0.32]{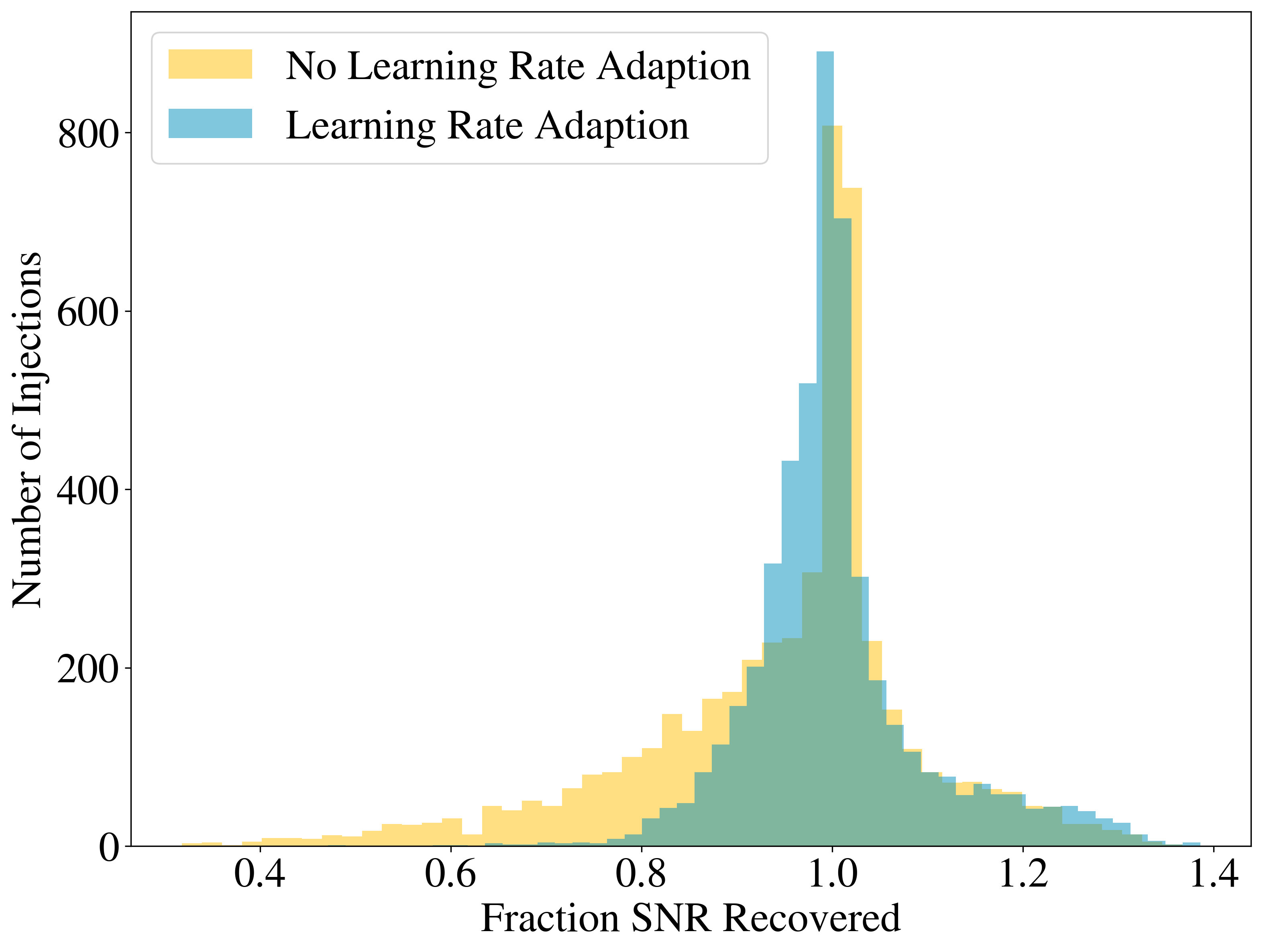}
    \caption{CMA-ES with learning rate adaption was explored and it was concluded that learning rate adaption enables more of the SNR to be recovered. This is supported by the literature that have previously shown that learning rate adaption and population restart is the best variation of CMA-ES for noisy problems  \cite{Nomura2023CMA-ESProblems, Hansen2009BenchmarkingTestbedb}.}
    \label{fig: LearningRate}
\end{figure}

\begin{figure}
    \centering
    \includegraphics[scale=0.35]{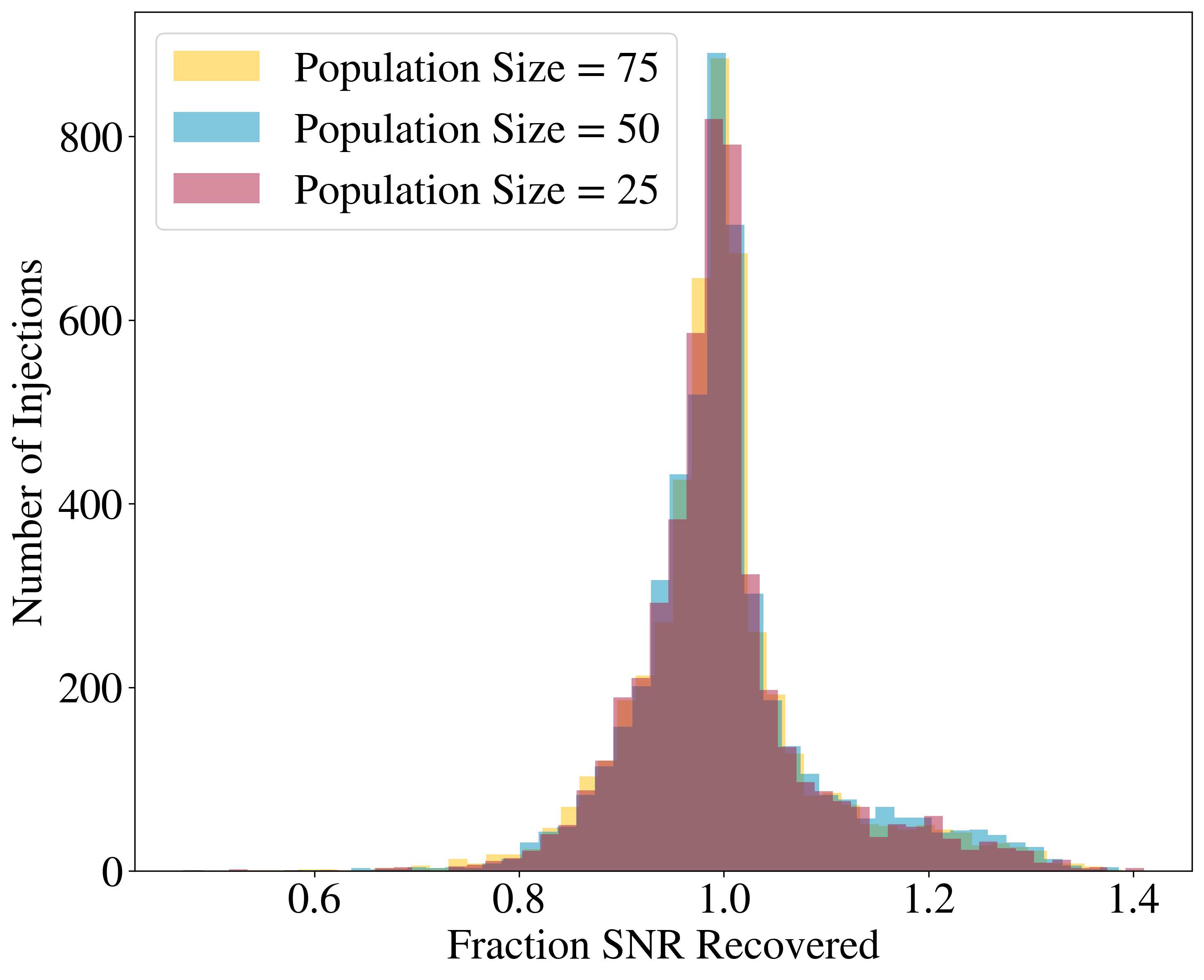}
    \caption{CMA-ES with population restart and learning rate adaption is used by GWtuna to recover the SNR once TPE has reached the SNR threshold of 6. Population size is required to be tuned when using population restart. The impact of population size on the fraction of SNR recovered was explored in this plot. It was concluded that population size of 50 was the most suitable because it was the fastest and recovered the SNR accurately.}
    \label{fig: Population size}
\end{figure}

In order to determine whether the software, algorithms and farmeworks presented in this paper were suitable for gravitational-wave detection, GWtuna was applied to a binary neutron star search. In this research we were interested in determining how well the combination of TPE and CMA-ES would be at recovering the SNR and template parameters of an injected binary neutron merger. Therefore, 5000 binary neutron star mergers with a variety of template parameters were injected into 512 seconds of coloured noise. The coloured noise was generated using a \textit{aligoO4low} PSD with a low frequency cut off of 18 Hz \cite{Ligo-Virgo-KAGRACollaboration2022LIGOT2000012-v2}. An IMRPhenomD waveform with a sampling rate of 2048 Hz and  a low frequency cut off of 20 Hz, was injected into the coloured noise. The reason for using an IMRPhenomD was that this was one of the few templates available on the JAX waveform library, \textit{Ripple} \cite{Edwards2023Ripple:Analysis}. The parameters of the injected template varied in the following ranges: $0.2736 \leq \lambda_{0} \leq  1.7073$, $0.1875 \leq \eta \leq 0.249999$, $-0.4 \leq \chi_{1,2} \leq 0.4$, $-40 s \leq tc \leq -20 s$, $6 \leq SNR \leq 20$. \textit{tc} is defined as the time of coalescence (measured in seconds) and its value determined where the template was injected into the coloured noise. The constraints on \textit{tc} were determined by the injection of the longest template into the coloured noise (with allowances for filter corruption). All parameters were sampled from a uniform distribution apart from SNR which was sampled from a custom distribution that focused on the lower SNR regions to replicate the gravitational waves observed in the universe. A single A100 GPU was used to gather the results presented in this paper and all code is available on Github at https://github.com/SusannaGreen/GWtuna. 

The mean time taken for a gravitational-wave signal to be identified and the parameters recovered by GWtuna is 49 seconds. It takes TPE and a stopping algorithm a median time of 1 second to identify a peak in SNR. Once the SNR threshold is surpassed, CMA-ES takes a median time of 48 seconds to recover the SNR and the parameters. Figure \ref{fig: TPE and CMAES Time} shows the time taken for TPE and CMA-ES to recover the peak in SNR for 5000 injections. A maximum of 1000 matched-filtering evaluations are used by TPE and 9000 matched-filter evaluations are used by CMA-ES. The number of matched-filter evaluations for TPE and CMA-ES was determined by trialling out different combinations and looking at the sensitivity of the GWtuna search. Therefore, less than 10,000 matched-filter evaluations are calculated in GWtuna when the stopping algorithm is used. It was determined that TPE would be stopped if the SNR did not improve in 500 matched-filter evaluations. This value was determined by balancing the accuracy of the GWtuna search and the computational time. GWtuna recovered 4972 out of 5000 injections as shown in Figure \ref{fig: SNR Actual and Predicted}. Out of the 28 injections GWtuna failed to recover, 2 had SNR values greater than 10. TPE failed to identify a peak in SNR over the range of SNR explored, see Figure \ref{fig: SNR Actual and Predicted}, and randomly throughout the parameter space, see Figure \ref{fig: Param recovery}. There is no discernible pattern of failure which indicates that the stopping algorithm may have been too constrained. Once TPE had identified a peak in SNR, CMA-ES is then used to recover the SNR and parameters by sampling the parameter space independent to TPE. $\lambda_{0}$, $\eta$, and aligned spins ($\chi_{1}$ and $\chi_{2}$) were then chosen to be searched over by TPE and CMA-ES, but other mass parameters were also tried. Figure \ref{fig: Mass Parameter} shows a comparison between the SNR recovered when $\lambda_{0}$ and $\eta$ or mass 1 and mass 2 or $\mathcal{M}_{c}$ and $\eta$ are searched over. CMA-ES recovered the $\lambda_{0}$ value the best compared to the other parameters, see Figure \ref{fig: Param recovery}. Furthermore, it is common occurrence that the spin values are not well-recovered. What appears to be a poor recovery of the parameters is because this is essentially a single detector problem and therefore many gravitational-wave search pipelines, including template bank searches, would struggle in recovering these parameters. 

There are many variations of CMA-ES as briefly discussed in \ref{GWtuna}. Initially in this research we tried warm-starting CMA-ES, CMA-ES learning-rate adaption, and CMA-ES with increasing/bi-directional population restart. Warm-starting CMA-ES used results from TPE to initialise CMA-ES however this required TPE to identify peaks in SNR accurately and therefore not mislead CMA-ES which is not feasible with a stopping algorithm. Population restart can be both increasing in population size or both increasing and decreasing in population size, otherwise known as bi-directional restart. In Figure \ref{fig: ipop or bipop} we showed that CMA-ES with increasing population restart recovered the SNR the best compared to CMA-ES with bi-directional population restart. As seen in Figure \ref{fig: LearningRate} without learning rate adaption the SNR is not recovered as well. For increasing population restart the size of the population is a tuneable parameter and therefore experiments were conducted to determine what value recovered the SNR the best in the shortest amount of time. It was concluded that a population size of 50 was able to recover the SNR accurately, see Figure \ref{fig: Population size}, in the fastest amount of time (median time of 50 seconds compared to a median time of approx. 80 seconds for the other population sizes). Population restart means that the CMA-ES will stop and then restart with 50 randomly sampled templates. In summary, we found that CMA-ES with increasing population restart and learning rate adaption recovered the SNR the best.
\label{Results}

\section{Discussion}
5000 binary neutron star mergers were injected into coloured noise to determine the feasibility of a GWtuna search. It was determined that TPE required less than than 1000 matched-filter trails to identify a peak in SNR within 1 second (median value) on a single A100 GPU. Whilst, CMA-ES required 9,000 matched-filter evaluations to recover the SNR and the template parameters in 48 seconds (median value). This could be accelerated by doing multiple matched-filter evaluations in parallel to hide GPU latency and the cost of the CMA-ES algorithm. 4972 out of the 5000 injections were successfully recovered. Out of the failed 28 injections, Figure \ref{fig: SNR Actual and Predicted} and Figure \ref{fig: Param recovery} show that TPE failed with no discernable pattern and therefore it was concluded that the stopping algorithm is too constrained. A stopping algorithm is used to curtail the TPE search if one of the following conditions are met: there is no consecutive improvement in SNR after 500 matched-filter evaluations or the SNR threshold is reached. If the SNR threshold is surpassed, then CMA-ES with population restart (see Figure \ref{fig: Population size} and Figure \ref{fig: ipop or bipop}) and learning rate adaption (see Figure \ref{fig: LearningRate}) is used to recover the SNR and the parameters. $\lambda_{0}$ and $\eta$, $\chi_{1}$, $\chi_{2}$ were chosen to be searched over by TPE and CMA-ES, but other mass parameters were also tried, shown in Figure \ref{fig: Mass Parameter}. The choice in parameters did have an impact on how well the the SNR was recovered and therefore for future research careful consideration should be given to the parameters used in TPE and CMA-ES. It is also worth mentioning that CMA-ES has been shown to be more efficient on larger parameter spaces and therefore future research will look into the feasibility of using CMA-ES to recover other parameters of gravitational-wave signals, other than the mass and spin parameters. This could mean that sky localisation could be recovered which would help in electromagnetic follow-up. However, brief investigations showed that there was a computational cost with regards to time taken in searching over more parameters.

The main aim of this paper is to showcase a non-template bank search prototype by introducing TPE and CMA-ES (and subsequently the family of black-box optimisation algorithms and evolution strategy algorithms) to the gravitational-wave community. Currently all matched-filtered gravitational-wave search pipelines used in the LVK Collaboration use template banks \cite{Adams2016, PyCBCSearch2016, Kovalam2022EarlySearch, Sakon2024TemplateKAGRA}. Template banks are inflexible which means that a number of matched-filter evaluations are always computed whether a gravitational wave is in the strain data or not. In contrast, GWtuna uses Bayesian optimisation to learn about each section of strain data and determine whether a gravitational-wave has been observed. Furthermore, stopping algorithms have been used to curtail optimisation if no gravitational-wave has been observed, which is a common scenario for binary neutron star mergers. At the moment template banks are not a huge problem because, for example, the PyCBC Live search pipeline takes on average 16 seconds to identify whether a gravitational wave has been observed (or not) using template banks \cite{Nitz2018RapidLive}. It is worth noting that in order for this speed to be achieved, PyCBC Live has been highly optimised and takes advantage of multi-processing (in contrast to GWtuna which is still a prototype). 

A GWtuna search pipeline would benefit the gravitational-wave community by providing a non-template bank search algorithm (meaning GWtuna is very generalisable) and as a consequence our understanding of a variety of gravitational-wave sources could be widened. In order for GWtuna to become a search pipeline, more thought is required to determine how it will be applied to actual interferometric data. A GWtuna pipeline would be required to mitigate non-Gaussian noise, otherwise known as glitches. Furthermore, verification steps would be needed to ensure that a gravitational wave has actually been observed. The ideas expressed in this research could be taken a step further and GWtuna could become a data-analysis pipeline, not just a search pipeline. CMA-ES is defined as a sampler and therefore, CMA-ES could be used to recover the parameters by sampling the posterior distribution rather than the SNR. As mentioned previously there would be a computational cost with sampling over more parameters however multi-processing on GPUs could be used to counteract this because CMA-ES intrinsically has a high degree of parallelism \cite{Nomura2021WarmOptimization}. Also, GWtuna is built on the JAX framework which contains functionalities that speed up a variety of processes in parameter estimation, including template generation, therefore GWtuna could become a fast data-analysis prototype. 

In the era of third generation \cite{Reitze2020CosmicLIGO, Punturo2010TheObservatory} and space \cite{Amaro-Seoane2017, Luo2016TianQin:Detector} gravitational-wave detectors, template banks are not a computationally feasible option. Recent publications for LISA have estimated template banks to be of the order of $\sim 10^{40}$ for extreme mass ratio inspirals (EMRIs) and stellar mass black hole mergers \cite{Moore2019AreLISA, Gair2004EventSources}. As the binary neutron star search is pushed into the lower frequencies (low frequency cut-off 7 Hz) in third generation gravitational-wave detectors, templates are of the order of 4000 seconds in length and therefore may cause memory issues \cite{Lenon2021EccentricExplorer}. There are also many problems that cannot be solved using template banks because the parameter spaces are either too large or not fully understood, such as precessing \cite{Brown2012NonspinningApproximation, Harry2014, Ajith2014EffectualSpins}, eccentric \cite{Brown2010EffectDetectors, Csizmadia2012GravitationalBinaries, Coughlin2015DetectabilityDetectors} and continuous wave searches \cite{Indik2017StochasticEvents}. In this paper, we have designed a GWtuna search for binary neutron star mergers in ground based detectors but this paper is about showcasing black box optimisation and evolutionary strategy algorithms to the gravitational-wave community. A new GWtuna search could be designed to facilitate searches for EMRIs in LISA but, how TPE and CMA-ES behave on other noise spectrums is unknown. Or a new GWtuna search could be designed for eccentric binary black hole mergers however, more research would be needed to determine how TPE and CMA-ES behave when other parameter spaces are searched over (only mass and spin parameters were explored in this research). The ideas expressed in this paper could be applied to the wider gravitational-wave community but a global effort would be required to determine how these algorithms would behave when applied to particular problems. 

\label{Discussion}

\section{Conclusions}
Binary neutron stars can be identified in less than one second without template banks because of GWtuna. GWtuna is a fast gravitational-wave search prototype built on Optuna (optimisation software library) and JAX (accelerator-orientated array computation library). In GWtuna, the SNR is recovered using a non-template bank search algorithm that combines TPE, a Bayesian optimisation algorithm, and CMA-ES. Both algorithms were originally designed for machine learning hyperparameter tuning but in this research we have shown that they also have the potential to be utilised in the gravitational wave community. It takes on average 1 second and less than 1000 matched-filter trails for a peak in SNR to be identified by TPE on a single A100 GPU. It then takes CMA-ES a further 50 seconds and 9,000 matched-filter evaluations for the mass and spin parameters to be recovered and SNR determined. If there is no gravitational-wave signal in the data, which is a common occurrence for binary neutron star searches, then a stopping algorithm will curtail the TPE search which prevents unnecessary calculations. This stopping algorithm will be initiated if one of the following conditions is fulfilled: the SNR has not improved in 500 matched-filter evaluations or the SNR threshold has been surpassed. A non template bank search, stopping algorithm and JAX functionalities (Just-in-time compilation which enables accelerated linear algebra and GPU compatibility utilised in this research) result in GWtuna being very computationally cheap. 

GWtuna is a search prototype and therefore no steps have been included to remove glitches or include data from all detectors. However, in the future a GWtuna search pipeline could be a possibility because GWtuna provides an alternative algorithm to template bank searches. Template banks are not computationally feasible for certain gravitational-wave sources, such as precessing compact binaries in ground based detectors and EMRIs in space based detectors. A GWtuna search pipeline could easily be generalised between gravitational-wave sources because all that is required is a waveform in JAX but research would be required to determine what parameters are searched over by TPE and CMA-ES. Other possibilities for GWtuna have been discussed in this paper, including using CMA-ES to sample around the maximum likelihood for all parameters therefore making GWtuna a data-analysis pipeline (not just a search pipeline). Even if a GWtuna pipeline is not feasible that the methods discussed in this research, specifically TPE and CMA-ES, and the family these algorithms originate from (hyperparameter optimization and evolution strategies) have the potential to revolutionise gravitational-wave astronomy and therefore the authors would encourage further research into these areas.

\label{Conclusions}

%\textit{Physical Review} style requires that the initial citation of
%figures or tables be in numerical order in text, so don't cite
%Fig.~\ref{fig:wide} until Fig.~\ref{fig:epsart} has been cited.

\begin{acknowledgments}
Susanna Green was supported by a STFC studentship and the University of Portsmouth. Andrew Lundgren acknowledges the support of UKRI through grants ST/V005715/1 and ST/Y004280/1. Numerical computations were done on the Sciama High Performance Compute (HPC) cluster which is supported by the ICG, SEPNet and the University of Portsmouth. This paper
has been assigned document number LIGO-P2400490.
\end{acknowledgments}

\bibliography{references}% Produces the bibliography via BibTeX.

\end{document}